\documentclass[prd,twocolumn,showpacs,preprintnumbers,floats,nofootinbib]{revtex4}
\usepackage{graphicx}
\usepackage{dcolumn}
\usepackage{bm}

\def\spose#1{\hbox to 0pt{#1\hss}}
\def\lta{\mathrel{\spose{\lower 3pt\hbox{$\mathchar"218$}}
     \raise 2.0pt\hbox{$\mathchar"13C$}}}
\def\gta{\mathrel{\spose{\lower 3pt\hbox{$\mathchar"218$}}
     \raise 2.0pt\hbox{$\mathchar"13E$}}}
\newcommand{\be}{\begin{equation}}
\newcommand{\en}{\end{equation}}
\newcommand{\bea}{\begin{eqnarray}}
\newcommand{\ena}{\end{eqnarray}}
\newcommand{\ex}{\mbox{e}}
\newcommand{\dd}{\mbox{d}}
\newcommand{\cL}{c_{_{\rm L}}}
\newcommand{\cT}{c_{_{\rm T}}}
\begin{document}

\title{Current-carrying cosmic string loops 3D simulation: \\
towards a reduction of the vorton excess problem.}

\author{Adriana Cordero-Cid}
\email{lcordero@sirio.ifuap.buap.mx}
\affiliation{Instituto de F\'\i sica, BUAP, A. P. J-48, 72570 Puebla,
Pue. M\'exico}

\author{Xavier Martin}
\email{xavier@fis.cinvestav.mx}
\affiliation{Dpto. de F\'{\i}sica, CINVESTAV-I.P.N. A.P. 14-74-, 07000
M\'exico, D.F., M\'exico}

\author{Patrick Peter}
\email{peter@iap.fr}
\affiliation{Institut d'Astrophysique de Paris, UPR 341, CNRS, 98bis
boulevard Arago, F-75014 Paris, France}
\date{January 11, 2002}

\begin{abstract}
The dynamical evolution of superconducting cosmic string loops
with specific equations of state describing timelike and
spacelike currents is studied numerically. This analysis extends
previous work in two directions: first it shows results coming
from a fully three dimensional simulation (as opposed to the two
dimensional case already studied), and it now includes fermionic
as well as bosonic currents. We confirm that in the case of
bosonic currents, shocks are formed in the magnetic regime and
kinks in the electric regime. For a loop endowed with a fermionic
current with zero-mode carriers, we show that only kinks form
along the string worldsheet, therefore making these loops
slightly more stable against charge carrier radiation, the likely
outcome of either shocks or kinks. All these combined effects tend
to reduce the number density of stable loops and contribute to
ease the vorton excess problem. As a bonus, these effects also
may provide new ways of producing high energy cosmic rays.
\end{abstract}


\maketitle


\section{Introduction}

Superconducting cosmic strings are topological defects~\cite{kibble}
that may be formed at a phase transition in the early
universe~\cite{witten,book}. Contrary to their usual counterpart known
as Goto-Nambu strings~\cite{GN}, these possess a nontrivial internal
structure because of conserved currents that can flow along
them~\cite{current,neutral,enon0}. The dynamics of superconducting
cosmic strings is, because of the current structure that has to be
taken into account, much more complicated. In order to describe their
evolution, the so-called elastic string formalism, thanks to which the
micro-structure can be integrated over to yield the macroscopic
behavior, was set up by Carter~\cite{formal}. The only requirement is
the knowledge of the equation of state relating the energy per unit
length to the string tension. This is equivalent to giving a single
Lagrangian function ${\cal L}(w)$, depending on a so-called state
parameter $w$, out of which the dynamical equations can be derived.

Various macroscopic Lagrangians for describing superconducting
cosmic strings have been proposed~\cite{models}, that correspond
to different assumptions about the internal structure. In
principle, the current along a cosmic string exists because some
extra degrees of freedoms (e.g. particles~\cite{witten}) are
trapped in the defect core. Up to now, there are essentially
three models that have been developed to yield analytic equations
of state. Those can be called respectively transonic, bosonic and
zero mode fermionic carrier models.

The transonic model describes for instance wiggly Goto-Nambu
strings~\cite{wiggles}. The trapped degrees of freedom in this case
are not particles, but rather transverse string excitations that play
the same role. It is self-dual, meaning that the equation of state
algebraic form does not depend on whether the current is timelike or
spacelike, and it is transonic because the velocities of transverse
and longitudinal perturbations happen to be equal.  Among its numerous
advantages, one finds that the transonic model can be exactly solved
in the case of flat space, and the string loop motion in this model is
always stable. It also describes the motion of an ordinary string in a
five-dimensional space-time, {\it \`a la} Kaluza-Klein~\cite{KK}. In
this last case, the state parameter represents the fifth coordinate of
the string location, seen by projection in the four dimensional
space-time.

Superconducting cosmic string may also involve actual particles,
the simplest case being that of fermions trapped in the defect
core in the form of zero modes, as originally proposed in 1985 by
Witten~\cite{witten} to introduce superconductivity. Under very
general circumstances, the fermion condensate can be
shown~\cite{fermions0} to rapidly reach a zero temperature
distribution, so that the integrated model does not depend on
anything else but the string characteristic mass scale, $m$ say.
The relevant equation of state in this case is then again
self-dual, with the energy per unit length and tension adding up
to a constant at all times.

Bosonic current-carrying cosmic strings are described by an equation
of state which is not self-dual. Such models require two different
masses~\cite{models}: the string scale $m$ and a mass characterizing
the current, $m_{*}$ say, usually taken to be that of the carrier
itself. Such models have been examined with full
details~\cite{neutral,enon0}. Even though a bosonic condensate can be
treated as a single classical field, contrary to it fermionic
counterpart, the corresponding strings look rather more involved than
the fermionic current-carrying ones. Indeed, it turned out that two
equations of state are required to describe bosonic current-carrying
cosmic strings, one for each of the available regime, namely spacelike
or timelike. What previous analysis revealed is that even in this
case, quantum effects should be taken into account, due to various
possible instabilities that could be triggered by the loops
evolution~\cite{2D}.

There are results on the evolution of superconducting string loops in
one and two dimensions~\cite{2D,1D} for those equations of state
corresponding to strings with bosonic currents. These works have shown
the appearance of various singular behaviors as the dynamical
evolution can drive parts of a given loop outside of the domain of
validity of the elastic string description. For instance, the loop can
develop shocks~\cite{2D,shock1} or fold on itself in complicated
shapes. The first question that could then be asked regarding these
effects is how much of these is due to the restriction to one or two
dimensions~? In other words, are these results real or do they merely
arise because of some projection effect. In the present paper, we have
accordingly extended previous studies to the simulation of a
superconducting cosmic string loop with bosonic currents in three
dimensions. We find that all the singular behaviors observed in two
dimensions generalize to three and cannot therefore be assigned to
some projection artifact.

Furthermore, we also investigated the behavior of a loop with
fermionic currents, using what was recently shown to be the
relevant equation of state for such
strings~\cite{fermions0,fermions}, i.e. the fixed trace model. In
this case, although dynamical evolution can also drive the loop
outside of the domain of validity of the elastic string
description and fold it in complicated shapes, shocks never
develop.  All these singular behaviors, both for fermionic and
bosonic carriers, are interesting because, they are expected to
lead to charge carrier emission, a mechanism that has many
cosmological consequences.

The dynamics of the string loops under consideration in this paper
demands to be applied to a network~\cite{network} of such strings in
order to decide on important issues such as vorton
formation~\cite{vortons1} and evolution~\cite{vortons2}. Another
related possibility concerns the up-to-now mysterious highest energy
cosmic ray events~\cite{uhecr}. These could be explained in terms of
topological defects such as superconducting cosmic strings, and indeed
there have been many such proposals~\cite{TDuhecr}. In the case at
hand, our conclusions seem to imply that much more particles are
released during the normal evolution of a current-carrying cosmic
string loop than during that of an ordinary loop. Assuming these
particles to have enough energy to contribute to the highest energy
cosmic ray, this paves the way to a new class of models that may not
suffer from the normalization constraint~\cite{gillkibble}.

The article is articulated as follows. In the next section we derive
the equations of motion in three dimensions, and express them in an
appropriate way for efficient numerical resolution. In section~III, we
present the equations of state for superconducting cosmic strings with
bosonic and fermionic currents, while the results of the simulation
are exhibited in section~IV. For a loop with a bosonic current, we
exemplify cases where the loop develops shocks, kinks or folds on
itself in much the same way as in the two dimensional case. The
analysis is then extended to superconducting loops with fermionic
zero-mode current-carriers.  In this case, we find that in some
instances, the loop shows regions of discontinuous curvature (kinks)
but we do not find shock waves as in the bosonic situation.  We also
observe that in most of the cases we have investigated, the dynamical
evolution of a loop, whether with bosonic or fermionic current, will
drive it out of the elastic regime. This is yet another confirmation
that the quantum (microscopic) effects that are taking place in a
cosmic string loop are almost always not negligible, and even more so
in the cosmological setting for which those effects might modify
drastically the model predictions~\cite{reliable}. We discuss this
particular point in the concluding section.

\section{Equations of motion}

For a current--carrying cosmic string, the equations of motion
can be expressed as the conservation of the stress energy tensor
and the equation of state. The stress energy tensor of a cosmic
string can be expressed in diagonal form as~\cite{formal}:
\begin{eqnarray}
T^{\mu\nu}=U u^{\mu}u^{\nu} - T v^{\mu} v^{\nu}, \end{eqnarray} where
$u^{\mu}$ and $v^{\mu}$ are the two orthogonal, respectively timelike
and spacelike unit eigenvectors\footnote{$u_\mu u^\mu = - v_\mu v^\mu
= -1$, $u_\mu v^\mu =0$}, which describe the string worldsheet, and
$U$ and $T$ are the two corresponding eigenvalues identified
respectively with the energy per unit length and tension of the
string. The equations of motion stem from the conservation of the
stress energy tensor, and they can be split into two pairs of
intrinsic and extrinsic dynamical equations. The intrinsic dynamical
equations are obtained by projection along the string worldsheet while
the extrinsic dynamical equations are obtained by projection
perpendicular to the string worldsheet. The intrinsic equations reduce
to two current conservation laws, one timelike
\begin{eqnarray} \bar{\nabla}_{\rho}( \nu u^{\rho} )= 0,
\end{eqnarray}
and the other spacelike
\begin{eqnarray}
\bar{\nabla}_{\rho}( \mu v^{\rho} )= 0, \end{eqnarray}
where
$\bar{\nabla}_{\rho}={\eta_\rho}^\sigma \nabla_\sigma$ is the
covariant derivative on the string worldsheet, obtained by projecting
the usual covariant derivative on the worldsheet with the projection
operator ${\eta_\rho}^\sigma=-u_{\rho}u^{\sigma}+v_{\rho}v^{\sigma}$, and
$\nu$ and $\mu$ are defined through
the relations
\begin{eqnarray} \mu \nu &=& U - T, \\ \mu &=& \frac{\dd
U}{\dd \nu},\end{eqnarray}
and can be understood, respectively, as a number density variable and
its associated effective mass variable or chemical potential.

The intrinsic equations can be equivalently re-expressed as
irrotationality equations, respectively as \begin{equation} \label{eiro1}
\epsilon^{\sigma \rho} \nabla_{\sigma} ( \nu v_{\rho})=0, \end{equation} and
\begin{equation} \epsilon^{\sigma \rho} \nabla_{\sigma} ( \mu
u_{\rho})=0,\label{eiro2} \end{equation} where $\epsilon ^{ \sigma \rho}$ is
the antisymmetric tangent element tensor of the string
worldsheet, given by
\begin{eqnarray}
\epsilon^{\sigma
\rho}=u^{\sigma}v^{\rho}-v^{\sigma}u^{\rho},\end{eqnarray}
in terms of the tangent vectors ${\bf u}$ and ${\bf v}$.

For a superconducting cosmic string, the conserved numbers
associated with the two conserved currents correspond to the
charged current trapped in the string and the winding number of
the string. When a loop is considered, these translate into two
integer-valued charges, $N$ and $Z$, say, once integration around
the loop is performed.

The extrinsic equations of motion can be written as\begin{eqnarray}
{\bot^{\mu}}_{\rho}(U u^{\nu} \nabla_{\nu} u ^{\rho} - T v^{\nu}
\nabla_{\nu} v^{\rho})=0, \label{eero} \end{eqnarray} where
${\bot^{\mu}}_{\nu}$, defined by  \begin{eqnarray} {\bot^{\mu}}_{\nu}=
{g^{\mu}}_{\nu}-{\eta^\mu}_\nu= {g^{\mu}}_{\nu}-
\epsilon^{\mu\rho}\epsilon_{\rho\nu}, \end{eqnarray} is the orthogonal
projection operator.

The necessary quantities for describing a cosmic string loop are
completed by the knowledge of two other important parameters,
namely the speeds of transverse and longitudinal perturbations
along the string, written respectively as $\cT$ and $\cL$, and
given by \begin{equation} \cT^2= \frac{T}{U}, \qquad \ \ \ \
\cL^2= -\frac{\dd T}{\dd U}.\label{cs}\end{equation} These
velocities are useful to explore the stability of a loop at
equilibrium~\cite{stab}. Furthermore, they are helpful since they
allow us to define what we call elastic regime: the string
dynamics is in the elastic regime if transverse and longitudinal
perturbations are stable. This indeed ensures that the dynamical
model does not break down. The velocities~(\ref{cs}) should also
be asked to both be less than unity in order to avoid
superluminal propagation, but this, in practice, never actually
happens so that it is not really a constraint. To summarize, we
require that, along the loop trajectories, we have
\begin{equation} 0 \leq c^2_{_{\rm T, L}}\leq 1, \end{equation}
constraints that are not always trivially satisfied.

We now have to fix the gauge to determine unambiguously the relevant
degrees of freedom. The dynamical equations are solved numerically
choosing as unknowns the string worldsheet coordinates $x^{\mu}(t,
\psi)$ as in~\cite{2D}, and as parameters $t$, the time coordinate,
and $\psi$, a spacelike curvilinear coordinate chosen as the potential
associated with the irrotationality equation (\ref{eiro1}). With this
choice, $u^{\rho}, v^{\rho}$ and $\nu$ are expressed as:
\begin{eqnarray}
u^{\rho}&=& \frac{\dot{x}^{\rho}}{\dot{\zeta}}, \\
v^{\rho}&=&\frac{1}{n \dot{\zeta}} (\beta \dot{x}^{\rho}
+ \dot{\zeta}^{2}x^{\prime\rho} ), \\
\nu&=&\frac{\dot{\zeta}}{n}
\end{eqnarray}
with dots and primes respectively denoting derivatives with
respect to $t$ and $\psi$ and we have adopted the following
definitions
\begin{eqnarray}
\dot{\zeta}&=& \sqrt{-\dot{x}^{\rho}\dot{x}_{\rho}}, \nonumber \\
\zeta^{\prime}&=& \sqrt{x^{\prime\rho}x^{\prime}_{\rho}}, \\
\beta&=& x^{\prime\rho} \dot{x}_{\rho} , \nonumber \\
n&=& \sqrt{\beta^{2}+ \zeta^{\prime 2}\dot{\zeta}^{2}}. \nonumber
\end{eqnarray}
The equation of motion (\ref{eiro1}) is automatically satisfied by
the choice of its associated potential $\psi$ as variable, and the
remaining equations of motion (\ref{eiro2}) and (\ref{eero})
become respectively
\begin{equation} {\bf v}  \cdot [(n^2 -
\beta^2 \cL^2) \ddot{\bf x} - \dot{\zeta}^4 \cL^2 {\bf
x}^{\prime\prime} - 2 \beta \dot{\zeta}^2 \cL^2 \dot{\bf
x}^{\prime}] = 0 , \end{equation} where ${\bf v}$ is the spatial
part of $v^{\rho}$, and
\begin{equation} {\bf w}_{\bot 1,2} \cdot [(n^2 - \beta^2 \cT^2) \ddot{\bf x}
- \dot{\zeta}^4 \cT^2 {\bf x}^{\prime\prime} - 2 \beta
\dot{\zeta}^2 \cT^2 \dot{\bf x}^{\prime} ] = 0,
\end{equation}
with ${\bf w}_{\bot 1,2}$ the spatial parts of two independent
quadrivectors orthogonal to the worldsheet, which we are choosing as
\begin{equation} w_{\bot 1}= \frac{1}{y^{\prime} \dot{z} -
\dot{y}z^{\prime}}(y^{\prime} \dot{z} - \dot{y}z^{\prime}, 0,
-z', y'),\end{equation} and \begin{equation} w_{\bot 2}=
\frac{1}{y^{'} \dot{z} - \dot{y}z^{\prime}}(0, y^{\prime} \dot{z}
- \dot{y}z^{\prime}, \dot{x}z'- \dot{z}x', \dot{y}x'-\dot{x}y').
\end{equation}
Then,  the equations of motion can be solved to find that $x(t,\psi),
y(t,\psi)$ and $z(t,\psi)$ satisfy \begin{eqnarray}
\ddot{x}&=& \frac{1}{n^{2}} ( x' F + G), \\
\ddot{y}&=& \frac{1}{n^{2}} ( y'F + I), \\
\ddot{z}&=& \frac{1}{n^{2}} ( z'F + K),
\end{eqnarray}
where the functions $F$, $G$, $I$ and $K$ are given by:
\begin{widetext}
\begin{eqnarray}
F&=& \frac{ \dot{\zeta}^{2} \cL^{2}} { n^{2} -
\beta^{2}\cL^{2}} \biggl[ \dot{\zeta}^{2} ( c_{1}x''+ c_{2}y''+
c_{3}z'') +2 \beta ( c_{1}\dot{x}'+ c_{2}\dot{y}'+ c_{3}\dot{z}')
\biggr] ,\\
G&=& \frac{ \dot{\zeta}^{2} \cT^{2}}{ n^{2} -
\beta^{2}\cT^{2}} \biggl\{ \dot{\zeta}^{2} \biggl[
(x'y''-x''y')c_{2} + (x'z''-x''z')c_{3} \biggr]
+2 \beta \biggl[ (x' \dot{y}'- \dot{x}' y')c_{2} + (x' \dot{z}'-
\dot{x}'z') c_{3} \biggr] \biggr\}, \\
I&=& \frac{ \dot{\zeta}^{2} \cT^{2}}{ n^{2} -
\beta^{2}\cT^{2}} \biggl\{ \dot{\zeta}^{2} \biggl[
(y''z'-z''y')c_{3}
+ (y''x'-x''y')c_{1} \biggr] +2 \beta \biggl[ (\dot{y}'z' -
y'\dot{z}')c_{3} + (\dot{y}'x'
- \dot{x}'y')c_{1} \biggr] \biggr\},\\
K&=& \frac{ \dot{\zeta}^{2} \cT^{2}}{ n^{2} -
\beta^{2}\cT^{2}} \biggl\{ \dot{\zeta}^{2} \biggl[
(y''z'+z''y')c_{2} + (z''x'-z'x'')c_{1} \biggr]
+2 \beta \biggl[ (\dot{y}'z'+ \dot{z}'y')c_{2} + (\dot{z}'x' - z'
\dot{x}')c_{1} \biggr] \biggr\} ,\end{eqnarray}
\end{widetext}
with

\begin{eqnarray}
c_{1}&\equiv& x' + \dot{y} \Delta_{3} + \dot{z} \Delta_{2} ,\nonumber
\\ c_{2}&\equiv& y' - \dot{x} \Delta_{1} + \dot{z} \Delta_{3} ,\\
c_{3}&\equiv& z' - \dot{x} \Delta_{2} - \dot{y} \Delta_{3} ,\nonumber
\end{eqnarray}
and
\begin{eqnarray}
\Delta_{1}&\equiv& y'\dot{x} - \dot{y}x' ,\nonumber \\
\Delta_{2}&\equiv& z'\dot{x} - \dot{z}x' ,\\
\Delta_{3}&\equiv& z'\dot{y} - \dot{z}y' .\nonumber
\end{eqnarray}
As a test of the deviations of the numerical approximation from the
exact solution it will be convenient to calculate the (conserved)
total energy and angular momentum of the system given, with our choice
of unknowns, by ($l$ being the string arc length)~\cite{2D}
\begin{equation} E_{_{\rm T}} \equiv \oint \dd l \,T^{00},
\end{equation}
which reads, explicitly
\begin{equation} E_{_{\rm T}} = \oint \dd \psi \frac{U[\nu(\psi)]}
{n\dot{\zeta}^2} (n^{2} - \beta^{2} \cT^{2} ),\end{equation}
and the angular momentum~\cite{formal}
\begin{equation}
J_{_{\rm T}}= Z N = \Psi \oint \dd \psi \frac{ U \beta
n}{\dot{\zeta}^2} (1 - \cT^{2}),
\end{equation}
where $Z$ is the particle quantum number, $N$ is the winding
number and $\Psi$ is the parametric length of the loop in the $\psi$
coordinate (which is conserved, actually being one of $N$ or $Z$).

We shall assume in what follows that the dynamics obtained by means of
the numerical solution is accurate as long as both these quantities
are conserved at least at the few $\%$ level (worst case during the
development of a shock) all along the trajectory.

\section{ Equation of state}

The equation of state describes the substructure of the
current--carrying cosmic string considered by relating its
internal parameters, for instance its energy density and tension.
There are three interesting current--carrying cosmic string
models: the transonic model and the bosonic and fermionic
superconducting models. The transonic model is used to describe
the macroscopic dynamical behavior of a wiggly ``Goto--Nambu''
cosmic string and carries a current associated with the
wiggles~\cite{wiggles}: it permits to neglect the dynamics of the
strings on small scales by integrating over these scales,
effectively raising the stress-energy tensor degeneracy, and
thereby to introduce an effective current. This model also
possesses the nice feature of being algebraically solvable in flat
space~\cite{integrability}, and thus can provide useful
analytical examples with which one can compare the accuracy of a
simulation code. The other models we shall deal with are used to
describe cosmic strings in which a current of particles (either
bosonic or fermionic) has condensed.

For superconducting strings with a bosonic current, the model
requires two mass scales, namely the cosmic string mass scale $m$,
which is the string forming symmetry breaking energy scale, and
$m_{*}$ which is the mass scale of the current carrier. In this
case, one can write down two different equations of state
depending on whether the conserved current along the string is
timelike or spacelike~\cite{neutral}. These two regimes of the
model are called respectively electric and magnetic in accordance
with the situation where the current is coupled to an
electromagnetic field~\cite{enon0}. Indeed, in this case, a
timelike current would induce an essentially electric field (in
the sense that, as ${\bf E}^2 > {\bf B}^2$, the magnetic
component can be framed away), while a spacelike current leads to
a magnetic-like field. Using a toy model which produces such
strings~\cite{witten,neutral}, the equation of state for U as a
function of T in the magnetic regime is~\cite{models}
\begin{eqnarray} U=m^{2}+\frac{m_{*}^{2}}{8}-\frac{m_{*}}{2}
\sqrt { \frac{m_{*}^{2}}{16} - m^{2} + T }
,\label{magneticUT}\end{eqnarray} and the transversal and
longitudinal speeds are given by
\begin{eqnarray}
\cT^{2}&=&\frac{1}{1+\nu_{*}^{2}}
\frac{ 2m^{2} (1+\nu_{*}^{2})^{2} -
\nu^{2}(1-\nu_{*}^{2})}{2m^{2}(1+\nu_{*}^{2})+\nu^{2}}, \\
\cL^{2}&=& \frac{1- 3\nu_{*}^{2}}{1+\nu_{*}^{2}},
\end{eqnarray}
where $\nu_{*}=\nu / m_{*}$. In this case, $\cL^{2}$ can become
negative so that the string may become unstable with respect to
longitudinal perturbations and thus leave the domain of
elasticity. In the electric regime, the toy model yields an
equation of state of the form \begin{equation} U= T + m_*^2\bigl[
\ex^{2(m^2 - T)/m_*^2} - 1 \bigr],\label{electricUT}
\end{equation}
and the two perturbations velocities are
\begin{equation}
\cT^{2}=\frac{\bigl[ 2(m^{2} / m^2_*) + \ln (1-X_{*})\bigr]
(1-X_{*})}{\bigl[ 2(m^{2} / m^2_*) + \ln (1-X_{*})\bigr]
(1-X_{*}) + 2X_{*}},\end{equation} where
\begin{equation} X_* = {2 \nu_*^2 +1-\sqrt{4\nu_*^2 +1}\over 2
\nu_*^2},\end{equation} and
\begin{equation}\cL^{2}= \frac{1}{\sqrt{1+4\nu_*^2}}.
\end{equation}
In this regime, it is clear that only $\cT^2$ may become
negative, at which point the string may become unstable with
respect to transverse perturbations.

The final model of current-carrying cosmic strings which we shall
consider here concerns the case in which fermions may be bound to
cosmic strings in the form of two lightlike currents of zero modes
propagating in opposite directions. The resulting total fermionic
current can then be timelike, spacelike or lightlike. Because the
fermion condensate tends to a zero temperature distribution, the
equation of state is self-dual, that is of the same form whether
the fermionic current is timelike or spacelike, given
by~\cite{fermions0,fermions}
\begin{equation} U+T= 2m^{2},\label{fermionic}\end{equation}
which gives \begin{equation} U = m^2 + {\nu^2\over 2},\qquad \ \ \
\ T = m^2 - {\nu^2\over 2},\end{equation} where $m$ is, as before,
the string scale. Here, the perturbation velocities are simply
\begin{eqnarray} \cT^{2}&=& \frac{2m^{2}}{U} - 1 = {2m^2-\nu^2\over 2m^2 +\nu^2}, \\
\cL^{2}&=&1. \end{eqnarray} In this case, the string can leave the
domain of elasticity only if the velocity of transverse perturbations
$\cT^2$ can become negative, i.e. if $|\nu | \geq \sqrt{2} m$. We
shall see that this indeed happens.

\section{Evolution}

We studied the evolution of loops in three dimensions after
having reproduced, as a check, exactly the two-dimensional
results of Ref.~\cite{2D}. Then, the evolution of loops in three
dimensions was investigated with the various superconducting
equations of state discussed above. The class of initial
configurations with which we have been concerned was obtained by
perturbing the circular equilibrium solution in the following way:
\begin{eqnarray}
x(t=0,\psi)&=& R \cos \left( \frac{\psi}{R \gamma \nu_{0}}\right) \\
\dot{x}(t=0,\psi)&=& -\frac{\cT}{1+\epsilon} \sin \left(
\frac{\psi}{R \gamma
\nu_{0}}\right) \\
y(t=0,\psi)&=& R e  \sin \left( \frac{\psi}{R \gamma \nu_{0}}\right) \\
\dot{y}(t=0,\psi)&=& \frac{\cT}{1+\epsilon} \cos \left(
\frac{\psi}{R \gamma
\nu_{0}}\right)\\
z(t=0, \psi)&=&A \cos\left(\frac{\ell \psi}{R \gamma \nu_{0}}\right) \\
\dot{z}(t=0, \psi)&=& - B \sin\left(\frac{\ell \psi}{R \gamma
\nu_{0}}\right) ,
\end{eqnarray}
where $\nu_{0}$ and
$\gamma\equiv(1-\dot{x}^{2}-\dot{y}^{2}-\dot{z}^{2}) ^{1/2}$ are
respectively the current and the Lorentz factor of the circular
equilibrium solution, $e$ is the ellipticity of the
configuration, $\psi$ is the spacelike coordinate, chosen to vary
from zero to $2\pi R \gamma \nu_{0}$ using the gauge freedom in
the worldsheet, $\epsilon$ is a dimensionless parameter which
measures the deviation in velocities from the circular
equilibrium state obtained when $e=1$ and $\epsilon=0$, and
$\ell$ is an integer phase parameter for the transverse $z$
motion. Note that the amplitude $A$ of the perturbation in the
transverse (with respect to the string loop plane) direction, and
that of the associated velocity $B$ are independent since those
perturbations have been shown~\cite{stab} to be absolutely stable.

For a given equation of state, the stability of the perturbed
circular equilibrium solution can easily be derived~\cite{stab}.
The primary purpose of this work is to find out the fate of the
unstable configurations. Before studying such configurations
within the framework of bosonic and fermionic superconducting
equations of state, we first checked the program against stable
ones with the transonic equation of state. In this case,
everything runs smoothly and there is nothing particular to note
about the loops dynamics. Thus, no evolution of transonic loops
is shown (see however Ref.~\cite{2D} for a discussion of these
configurations; the three dimensional simulation now presented
merely confirms this reference results as we were able to check
that they were not artifacts of the two dimensional projection).

More interesting is the case of the bosonic current where singular
dynamics where observed in two dimensions~\cite{2D}. The analysis
in this case is presented next. It is then followed by an
investigation of the fermionic current case which is completely
new.

\subsection{Bosonic Currents}

For superconducting strings with bosonic currents, we confirm that all
the singular behaviors observed in two dimensions, i.e., shocks in the
magnetic regime, kinks in the electric regime, folding loops and exit
from the elastic regime in general, persist in three dimensions~; they
are not projection artifacts and represent the real physical time
evolution of superconducting cosmic string loops. In the following, we
show examples that illustrate each case in turn.

\begin{figure}[t]
\includegraphics*[width=8.5cm]{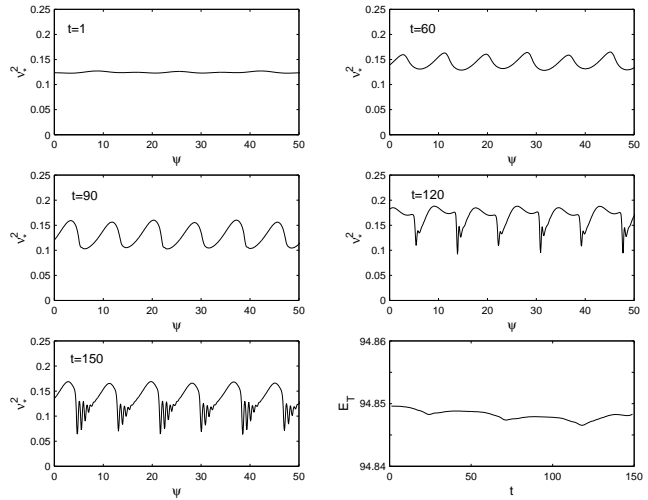}
\caption{Time evolution of the state parameter $\nu_{*}^{2}$ along the
string worldsheet for an elliptic configuration of a loop with a
bosonic current in the magnetic regime, having large ellipticity
$e=0.6$ and velocity parameter $\epsilon=1$, and with a phase
parameter $\ell=3$ (and $A=3$, $B=-0.2$). In the last panel the
variation of the energy shows two important things. On the one hand
the small peaks in the graph are an evidence that the loop size
decreases because the initial velocity has been reduced. On the other
hand, nearly at the end of the simulation, the rare variation of the
energy gives a signal of the shocks produced along the loop
string. The relative variation of the total energy, shown on the last
panel, is still rather small even as the simulation reaches its very
end. The situation is similar for the angular momentum, and the
simulation eventually breaks down.
\label{shock}}
\end{figure}

A shock wave may be formed at points along the string worldsheet,
and according to previous results \cite{2D,shock1} we further
confirm that a shock can only take place in the magnetic regime.
An example of such a shock in three dimensions, using equation of
state (\ref{magneticUT}) is shown in Fig.~\ref{shock} on which is
also represented the space variations of the state parameter
$\nu_*= \nu / m$ for different instants in time for a transversely
perturbed loop with $\ell=3$~; note that such a
loop, from the transverse perturbation point of view, should be
strictly stable. In this case, the geometric evolution does not
reveal any visible change in the loop, but the time evolution of
the state parameter $\nu_{*}^{2}$ as a function of the $\psi$
parameter shows the development of shock fronts along the string
worldsheet. At these points the current in the string becomes
discontinuous and the code cannot handle the evolution anymore, as
shown for instance by the fact that the total energy is no longer
conserved. Although such transverse modes in principle would have
been expected to be stable, it turns out, with the values of the
parameters chosen, that nonlinear effects rapidly become
dominant, effectively coupling the transverse mode to the
longitudinal one. That this really occurs is made clear by the
fact that the number of shock fronts generated during this
evolution is $2\ell$, thus directly related to the phase
parameter $\ell$.

\begin{figure} \includegraphics*[width=8.5cm]{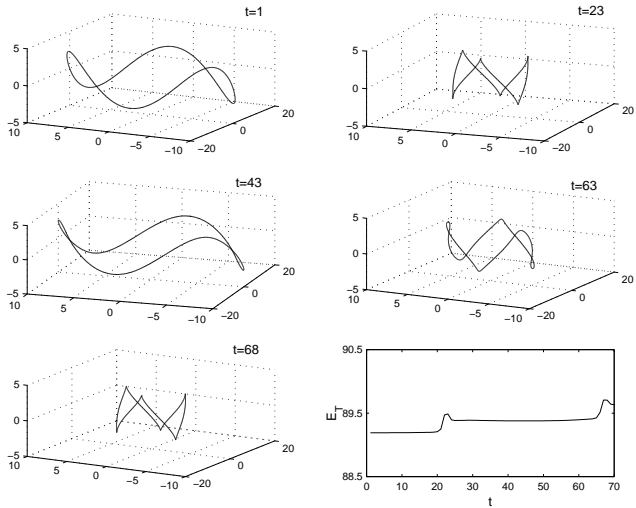}
\caption{Configuration evolution in the electric case with bosonic
currents with the same ellipticity as in Fig.~\ref{shock},
$\epsilon=0.1$ and $A=3$ ($B=0$). One clearly sees the development of
six kinks corresponding to twice the phase parameter $\ell=3$. In this
case, the peaks seen in the energy are formed at the same time as the
kinks are present.
\label{kink1}} \end{figure}

\begin{figure}
\includegraphics*[width=8.5cm]{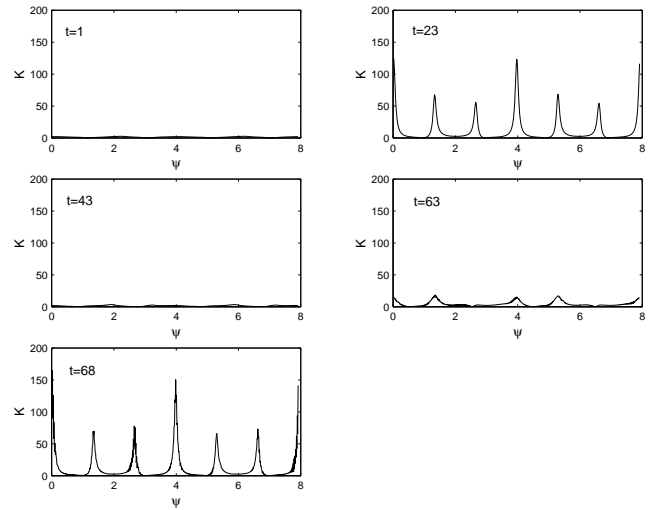}
\caption{Invariant magnitude $K=\sqrt{K^2}$ of the curvature, from
Eq.~(\ref{K2}), for the configuration shown in
Fig.~\ref{kink1}. \label{kink2}}
\end{figure}

Another important effect that may take place along a
superconducting loop is the appearance of kinks. These are
regions where the curvature of the string becomes discontinuous.
We found that giving an extra dimension for the loop to evolve
into did not suppress the appearance of kinks. Fig.~\ref{kink1}
shows the evolution of a loop in the electric regime, and
therefore using the equation of state given by
Eq.~(\ref{electricUT}). In this case again, the simulation shows
that $2\ell$ kinks are being formed at points along the loop. To
illustrate this point further, we plotted in Fig.~\ref{kink2} the
curvature of the loop as a function of the spacelike string
internal coordinate $\psi$. This is, in general, given
by~\cite{formal}
\begin{equation} K^\mu = \eta^{\rho \nu} K_{\rho
\nu}^{\mu},\end{equation}
i.e., in the case at hand,
\begin{equation} K^{2} = (w_{\bot 1}^{\rho} K_{\rho})^{2} + (w_{\bot
2}^{\rho} K_{\rho})^{2}, \label{K2}\end{equation} where
\begin{widetext}
\begin{eqnarray}
w_{\bot 1}^{\rho} K_{\rho}&=& \frac{1}{\Delta_{3} n^{2}
\dot{\zeta}^{2}} \left[ \left(n^{2}  - \beta^{2}\right)
\left(\ddot{y} z' - \ddot{z} y'\right) + 2 \beta \dot{\zeta}^{2}
\left(y' \dot{z'} - z'\dot{y'}\right)  +  \dot{\zeta}^{4}
\left( y' z'' - z' y''\right) \right], \\
w_{\bot 2}^{\rho} K_{\rho}&=& \frac{1}{\Delta_{3} n^{2}
\dot{\zeta}^{2}} \left[ (n^{2} - \beta^{2})(\Delta_{3} \ddot{x} -
\Delta_{2} \ddot{y} + \Delta_{1} \ddot{z})-2 \beta \dot{\zeta}^{2}
( \Delta_{3} \dot{z'} - \Delta_{2} \dot{y'} + \Delta_{1} \dot{z'})
- \dot{\zeta}^{4} ( \Delta_{3} x'' - \Delta_{2} y'' + \Delta_{1}
z'') \right] \!\! .
\end{eqnarray}
\end{widetext}
It is clear in the figure that where kinks appear, large
variations of the curvature are developed, showing that the kinks
are not gauge artifacts but really occur. As expected, the
location of the kinks corresponds to the points with large
curvature. In general the  effects found in the two dimensional
planar case are still present in three dimensions. They are
however usually affected (or even initiated) by the transverse
perturbation in $z$.

\begin{figure}
\includegraphics*[width=8.5cm]{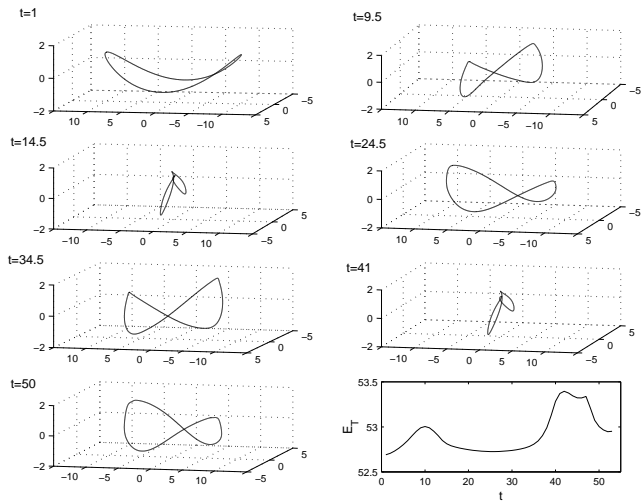}
\caption{A typical string loop configuration showing points where
intercommutation should take place, with $\ell=2$, ellipticity
$e=0.3$, $A=1$, $B=0.1$, and $\nu_{ * 0}^{2}=0.1$. At these points,
two loops should be formed and the subsequent evolution given by the
code is no longer physically relevant. \label{fold}}
\end{figure}

In some cases, the string loop shows points where
intercommutation, which is not taken into account in our code,
should be occurring. A systematic analysis shows that a loop in
three dimensions tends to fold on itself in any regime, magnetic
or electric. This is similar to what happens to Goto-Nambu loops
for which it was even argued that this intercommuting process
could be the leading process for a string network to loose
energy~\cite{reliable,topoconj}. The conclusions that can be drawn
from this fact for a superconducting string network are therefore
the same, and will be discussed later. In Fig.~\ref{fold} is shown
an example of such a missed intercommutation for a loop in the
electric case.

\begin{figure}
\includegraphics*[width=8.5cm]{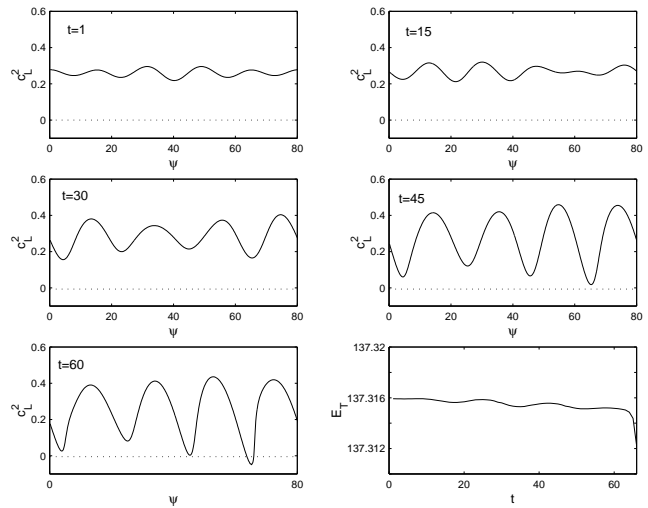}
\caption{Time evolution of the squared perturbation velocity,
$\cL^{2}$, in the magnetic regime for a bosonic superconducting
loop. The dynamics drives the loop out of the elastic regime, as seen
on the last evolution panel, in which one finds a region having $\cL^2
\leq 0$. The last panel shows the total energy integrated along the
loop as a function of time: its constancy up to the point where
$\cL^2$ becomes negative reveals that nothing particular took place
and the physical description can be trusted. Past this point, the
energy is very badly conserved and the program, as expected, has
difficulties handling the problem. This figure is for $A=2$ and
$B=0$. \label{outbo}}
\end{figure}

Another possibility that is open for a loop to leave the region
where the elastic description is valid is to locally have one of
its squared perturbation velocity, $\cL^{2}$ in the magnetic
regime, or $\cT^{2}$ in the electric regime, becoming negative. In
the latter case, the tension reaches negative values and the
string actually behaves like a spring. In Fig.~\ref{outbo} is
shown the evolution of the longitudinal perturbation velocity as a
function of the spacelike parameter along the string loop at
different instants of time in a typical example for which the
dynamical evolution of the loop drives it out of the elastic
regime when $\cL^{2}$ becomes less than zero. Whatever happens
past this time is presumably not accounted for correctly by the
program which, being an effective macroscopic description, does
not recognize the instability at the microscopic level.

All these results are reminiscent of what had previously been
obtained in the two dimensional case~\cite{2D}. There now is no
more ambiguity about them being projection artifacts. The
conclusions derived in the 2D case therefore still apply.

\subsection{Fermionic Current}

\begin{figure}
\includegraphics*[width=8.5cm]{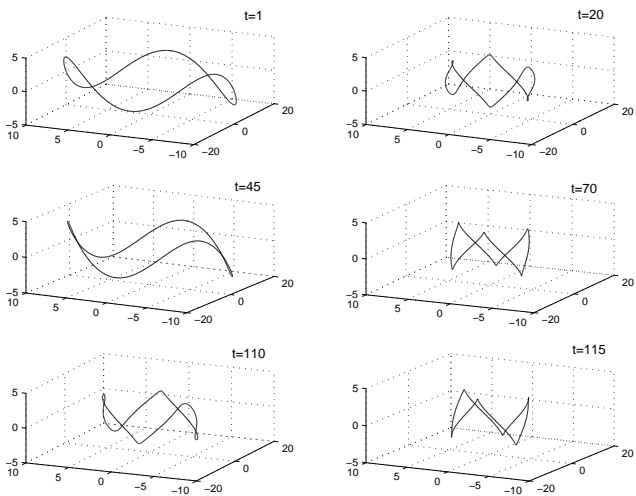}
\caption{Configuration evolution for fermionic equation of state.  In
this case, the shape of the loop changes until kinks are formed along
the string, at which point the elastic string description ceases to be
sufficient. Quasi kinks are formed first, that are subsequently
rounded off, until the actual kinks really form. Here, and in the
following figure, the parameters $A=3$ and $B=0.2$ have been
used.\label{kfl22}}
\end{figure}

\begin{figure}
\includegraphics*[width=8.5cm]{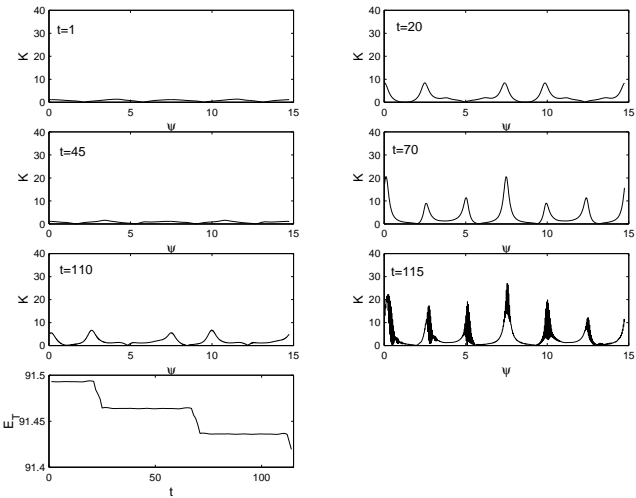}
\caption{Time evolution of curvature against the point labeling
function $\psi$ for the configuration of the Fig.~\ref{kfl22}. The
curvature in this example varies almost by two orders of
magnitude. As in the bosonic case, the variations of the energy
give a signal of the kinks formation. \label{ccf}}
\end{figure}

In this section, we consider current--carrying cosmic strings with
fermionic currents, obeying the equation of state
(\ref{fermionic}). The behavior in this case follows somewhat that
observed in the bosonic case: the loop can fold on itself and leave
the elastic domain. We shall exemplify both cases in turn, but first
let us remark the important following point.  Since in the fermionic
case the equation of state is self-dual, there cannot be any
qualitative difference between the magnetic and the electric regimes
at the level of the equations of motion.  However, the two
perturbation velocities are not equal and the string may become
unstable only with respect to transversal perturbations. As a result,
the dynamical evolution of the string can lead it to develop only
kinks\footnote{Note that since shocks can only form in the magnetic
regime and as in this self-dual situation, the electric and magnetic
regime are undistinguishable, shocks cannot form in the fermionic
current case.}, although in this case we cannot say that the effect is
exclusive of the electric or magnetic regime. In Figs.~\ref{kfl22} and
\ref{ccf}, we show the time evolution and the curvature (snapshots at
different times) for a loop with fermionic current, again with phase
parameter $\ell=3$ and initial value of the state parameter set as
$\nu_{* 0 }^{2}=0.1$. The appearance of kinks is seen as the curvature
of the string becomes discontinuous. When the kink is formed the total
energy is no longer conserved and the simulation ends.

Similarly to the case of cosmic string loops endowed with bosonic
currents, loops with fermionic currents can fold on themselves,
exhibiting points where intercommutation should take place, and
with very much the same rate of occurrence in the evolution.
Self-intercommutation for loops loosing their energy again seems
to be in this case a favored mode.

\begin{figure}[h]
\includegraphics*[width=8.5cm]{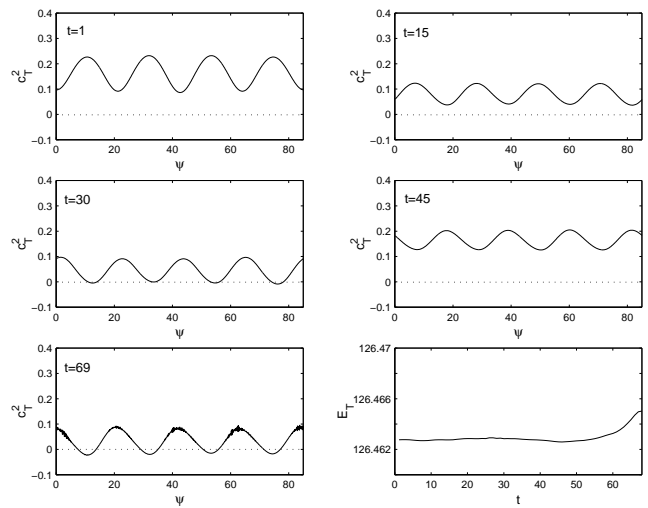}
\caption{Time evolution of the speed of transversal perturbations for
a cosmic string loop with fermionic currents (with $A=2$ and
$B=-0.2$). The dynamical evolution of the loop drives it out or the
elastic regime, where $\cT^2\leq 0$. \label{sfe}}
\end{figure}

The string loop can also leave the domain of validity of the
elastic formalism when $\cT^{2}$ becomes negative. Fig.~\ref{sfe}
depicts the time evolution of the transverse velocity $\cT^2$ up
to the point where it vanishes. The velocity here is shown as a
function of the spacelike string parameter $\psi$. The same
remarks as for bosonic strings apply, in particular concerning
the non conservation of the total energy whenever the elastic
regime is left. Note here the appearance of a numerical
instability that propagates along the string and that is seen in
the form of (unphysical) oscillations.

\section{Discussions and Conclusions}

This work is the direct follow-up of previously discussed two
dimensional string loop simulations~\cite{2D}. It extends this
old work in two directions. First it now provides results about
the dynamics of such string loops in three dimensions, thereby
answering the question as to whether the observed effects were
actually physical or mere projection artifacts. This new
simulation clearly shows that indeed the effects presented in
Ref.~\cite{2D} are physical. Another extension of Ref.~\cite{2D}
that is done here concerns the equation of state, which can now
describe fermionic~\cite{fermions0} currents as well.

In the case of strings endowed with spacelike bosonic currents, it
was shown that most of the loops develop shock. These shocks are
seen as discontinuities in the current and occur because the
squared longitudinal perturbation velocity $\cL^2$ can become
negative. In the electric regime of timelike currents, it is the
squared transverse perturbation velocity $\cT^2$ that can vanish
and reverse its sign. As it turns out, this implies that kinks,
namely regions of discontinuous curvature, develop.

Fermionic strings, being described by an equation of state for
which only the squared transverse perturbation velocity can
become negative, are subject only to the possibility of building
kinks, and this in a way that is independent of the spacelike or
timelike nature of the underlying current since the relevant
equation of state is self-dual.

In both cases, fermionic and bosonic currents, and for whatever
kind of current, the loops tend to fold on themselves, generating
contact points where intercommutation should occur. The almost
systematic occurrence of such configuration leads us to conclude
that one may divide superconducting string loops into essentially
two species: those loops which tend to divide into smaller loops,
and the other ones that are identified with the proto-vortons of
other authors~\cite{vortons2}. Our results tend to indicate that
the latter category is much less populated than the former, as
was already postulated in earlier investigations~\cite{vortons2}.
It also implies that these intercommutations will be, in the case
of superconducting cosmic strings, much more frequent than in the
case of ordinary strings.

Now when a string loop self-intercommutes, the topological stability
that confines the Higgs and other particles inside the defect is
raised and some of these particles are expelled away.  Although stable
while contained in the defects, these particles are often unstable. In
particular, strings are expected to form at the Grand Unified (GUT)
scale, and the relevant Higgs and gauge fields to decay therefore
almost instantaneously by cosmological standards. These particles will
then initiate showers of lower mass ones dividing the original
energy. This is the mechanism that is at work in most Top-Down
scenarios~\cite{topdown} of Ultra High Energy Cosmic Rays (UHECR)
whose origin is still mysterious~\cite{uhecr}. Ordinary cosmic strings
do not produce much of these particles, because intercommutations are
not that frequent in Goto-Nambu strings since most of the energy of
the network in that case is expected to be lost through gravitational
radiation~\cite{book}. In fact~\cite{gillkibble}, the resulting flux
is found a disappointing ten orders of magnitude below the observed
one. In the case of current-carrying strings, loops have a higher
probability of self-intercommuting, so the mechanism is enhanced
accordingly.

The proto-vorton case now takes us to the rare configurations
that do not intercommute frequently. Note at this point that the
scarcity of these states is not enough to ensure the vorton
excess problem to be altogether alleviated. Indeed, for GUT scale
vortons, the ratio between proto-vortons and doomed loops should
already be much less than $10^{-20}$ in order for the present-day
vortons not to dominate the Universe~\cite{vortons2} if the
resulting vortons are stable over cosmologically relevant
timescales.

Gathering all the results obtained over the recent years, we now
find the following: for a realistic underlying microscopic model,
the equation of state is such that at least one of the
perturbation propagation speed can become imaginary, leading to
possible instabilities. The main result of the present paper is
to show that these instabilities will systematically develop in
every loop as it evolves. As a result, the elastic treatment
turns out to be valid for limited periods of time and one must
resort to the full microscopic description in between.

The elastic description breaks down in two possible ways, namely
through the formation of kinks or shocks. In both cases, the
meaning is that the microstructure should be taken into account.
In particular, close to these points, it is no longer appropriate
to consider the string as effectively two dimensional and we must
consider the effects that are due to the finite thickness.

First of all, kinks cannot exist if the string is a tube instead
of a Dirac line as it would be impossible for instance to match
the Higgs field and its derivatives, as is however required by
the underlying field theory, at the points where the curvature is
discontinuous. Therefore, the string must somehow find a way of
smoothing, which means in that context loosing energy. The only
non kinematical (in the sense of necessarily breaking the elastic
description) way of realizing that is through particle emission,
and we are led to conclude that, as in the case of
intercommutation, but for completely different reasons, some
particles (current carriers, but also Higgs and gauge fields)
will be radiated away as the kinks develop. Note also that the
current-carrier is indeed likely to get out of the string since,
as the string curvature increases, the carrier momentum acquires
an orthogonal-to-the-worldsheet component thanks to which the
wavefunction is spread over larger distances, thereby effectively
increasing the probability for the particle to tunnel out. This
mechanism can ease the vorton excess problem while in the same
time providing more efficiency for UHECR production through
Top-Down scenarios.

Shocks develop only for bosonic spacelike currents. In this case,
the state parameter $\nu$, which would end up being discontinuous
were the shock actually occurring, is well represented by the
collective momentum of the trapped
particles~\cite{neutral,enon0}, this momentum being always less
than the rest mass of the carrier particles themselves in order
for the confinement to take place. As we argued before~\cite{2D},
it is rather dubious that actual shock could form in cosmic
strings, as the shocks are always accompanied, either followed or
preceded, by a transient region in which $\cL^2\leq 0$. However,
in this region, the momentum of the carrier is higher or at least
of the same order as its rest mass. Therefore, it is energetically
favored to some of the trapped particles to move out of the
string.

In all the situations which we have studied, we found that the
expected dominant effect was particle radiation. We are therefore
led to the conclusion that an elastic treatment of cosmic strings
is not completely appropriate. It should be supplemented with
some semi-classical, non conservative, forces. The dissipative
effect then introduced can serve to ease the vorton excess
problem, while providing new interesting ways to produce UHECR.

\acknowledgments We should like to thank B. Carter and C.~Ringeval for
many fruitful and enlightening discussions as well as A. Zepeda for
support during the preparation of this paper. L.~A.~C. acknowledges
the support received from CONACyT.

\end{document}